\documentclass{optica-article}

\journal{opticajournal} 

\articletype{Research Article} 

\usepackage{lineno}

\usepackage{mathtools}

\usepackage{cleveref}
\usepackage{xparse} 
\usepackage[normalem]{ulem} 
\usepackage{soul} 
\usepackage[separate-uncertainty = true]{siunitx}
\usepackage{diagbox}

\usepackage{tikz}
\usetikzlibrary{arrows,shadows,positioning}

\tikzset{
  frame/.style={
    rectangle, draw,
    text width=6em, text centered,
    minimum height=3em,drop shadow,fill=white,
    rounded corners,
  },
  line/.style={
    draw, -latex',
  },
  mynode/.style={
    draw, left=2pt, align=center,
  }
}

\newcommand{\myfraction}{0.75}

\begin{document}

\title{Image-based wavefront correction using model-free Reinforcement Learning}

\author{Yann Gutierrez\authormark{1,2,3,*}, Johan Mazoyer\authormark{1}, Laurent M.\ Mugnier\authormark{3}, Olivier Herscovici-Schiller\authormark{2}, and Baptiste Abeloos\authormark{2}}
\address{\authormark{1}LESIA, Observatoire de Paris, Université PSL, Sorbonne Université, Université Paris Cité, CNRS, 5 place Jules Janssen, 92195 Meudon, France \\
\authormark{2}DTIS, ONERA, Université Paris Saclay, 91123 Palaiseau, France \\
\authormark{3}DOTA, ONERA, Université Paris Saclay, BP 72, 92322 Châtillon cedex, France \\
\email{\authormark{*}yann.gutierrez@obspm.fr, yann.gutierrez@onera.fr}}

\begin{abstract*}
Optical aberrations prevent telescopes from reaching their theoretical diffraction limit. Once estimated, these aberrations can be compensated for using deformable mirrors in a closed loop. Focal plane wavefront sensing enables the estimation of the aberrations on the complete optical path, directly from the images taken by the scientific sensor. However, current focal plane wavefront sensing methods rely on physical models whose inaccuracies may limit the overall performance of the correction. The aim of this study is to develop a data-driven method using model-free reinforcement learning to automatically perform the estimation and correction of the aberrations, using only phase diversity images acquired around the focal plane as inputs. We formulate the correction problem within the framework of reinforcement learning and train an agent on simulated data. We show that the method is able to reliably learn an efficient control strategy for various realistic conditions. Our method also demonstrates robustness to a wide range of noise levels.
\end{abstract*}

\section{Introduction}
\label{sec:intro}

\subsection{Context}

Space telescopes allow astronomers to study faint, distant objects and capture details obscured by the Earth's atmosphere, offering a complementary view to ground-based observations. Their performance is limited by intrinsic aberrations, which stem from imperfections in the design and slowly evolving thermo-mechanical constraints. These aberrations are quasi-static, and may significantly degrade scientific images. Active optics methods correct the aberrations using a deformable mirror (DM) operating in a closed loop to recover and maintain an angular resolution close to the diffraction limit. The DM commands are derived from estimations of the wavefront aberrations, which pupil plane wavefront sensors (WFS) cannot measure all the way down to the imaging focal plane. In contrast, focal plane wavefront sensing (FPWFS) methods use a physical model of the instrument in conjunction with the focal plane images to estimate the aberrations without requiring any additional optics besides the imaging sensor~\cite{Mugnier-l-06a}. 

Early FPWFS methods relied on iterative projection-based algorithms that find the aberrations most compatible with known constraints in the pupil and focal plane \cite{gerchberg_practical_1972, Gonsalves:76, Fienup:82}. Conventional ``phase retrieval'' algorithms suffer from a sign ambiguity in the estimation, corresponding to a non-unique response in the focal plane for some aberrations when using a single image. To address this limitation, a method known as Phase Diversity was developed \cite{gonsalves1982, Mugnier-l-06a} which introduces an additional image with a known phase variation alongside the scientific data.
The second image, often called the diversity image, allows one to lift the sign ambiguity. 
While the diversity image takes some resources away from the science image, it is not necessarily an important overhead. Both focused and defocused images can be acquired simultaneously on the same sensor, using, \emph{e.g.}, a small periscope with a beamsplitter, so that the temporal evolution of aberrations is not a problem.  The sensor can also use a dedicated spectral channel if appropriate.
However, this technique comes at a high computational cost, as the problem is non-linear, making it difficult for real-time applications unless linearized \cite{Mocoeur-a-09b}. Additionally, model inaccuracies can limit correction precision to a few nm. 

In parallel, deep learning has emerged as a powerful tool for FPWFS, offering potential improvements in speed and complexity compared to traditional methods. Early studies laid the groundwork, demonstrating the ability of neural networks to estimate low-order and static aberrations in both ground-based \cite{angel_adaptive_1990, sandler_use_1991}, and space-based telescopes \cite{barrett_artificial_1993}, while later developments in convolutional neural networks (CNNs) have significantly expanded the capabilities.
CNNs have been used to estimate a range of Zernike coefficients and even directly reconstruct the pupil plane phase \cite{guo_improved_2019, wang_deep_2021}, with varying degrees of precision. These approaches often leverage phase diversity images \cite{quesnel_deep_2020, orbandexivry_focal_2021, wu_sub-millisecond_2020, andersen_neural_2019, andersen_image-based_2020}. Other approaches have been explored, using a single image 
\cite{paine_machine_2018, tian_dnn-based_2019, nishizaki_deep_2019, khorin_neural_2021}, or extracted features such as Chebyshev moments \cite{ju_feature-based_2018} or frequency features \cite{xin_object-independent_2019, naik_convolutional_2020}, as inputs. 

Despite these advances, a key challenge lies in the reliance on large labeled datasets for training, which can be difficult and expensive to obtain from real observations. While simulations offer an easier path to data generation, they introduce dependence on the physical models used, potentially limiting real-world precision.
To circumvent this limitation, some authors \cite{quesnel_simulator-based_2022, xu_self-supervised_2022} explored the potential of unsupervised learning by using autoencoders to learn Zernike coefficients directly from phase diversity images without requiring labels. 

However, the relation between the DM commands and the resulting shape applied to the surface of the DM may also be non-linear depending on the technology, and correction algorithms usually predict this interaction using another physical model. While this non-linearity is not a critical issue for ground-based AO systems operating in closed-loop, it can become problematic for some high-precision applications. Hence, regardless of the quality of the estimation, the correction precision can still suffer from inaccuracies in the physical model linking DM adjustments to their actual impact on the image, potentially resulting in a discrepancy between the estimated phase and the phase actually corrected. 

A sub-field of deep learning that has shown promise for learning to perform complex tasks without requiring supervision is deep reinforcement learning. This technique combines deep learning with reinforcement learning (RL), where an agent learns a control task by interacting with its environment. Through trial and error, the system learns to maximize a reward signal by taking actions and observing their outcomes. This reward function, calculated directly from observations, guides the system towards effective strategies without requiring accurate models. This data-driven approach allows training on real-world data without manual labeling. Furthermore, by directly learning actions from observations, RL can potentially perform the estimation and correction steps simultaneously, potentially achieving higher accuracy, as errors in estimation can be amplified by the correction when both steps are performed sequentially. In this work, we study the possibility of using model-free RL for focal plane wavefront control from phase diversity images of a point source. 

\subsection{Related work}
\label{subsec:sota}

The potential of using RL for wavefront control was first discussed as a promising data-driven correction method for High Contrast Imaging \cite{sun_identification_2018, radhakrishnan_optimization_2018}. Since then, research in this field has flourished, with various approaches emerging. 
While the work presented herein focuses on applying RL to focal-plane wavefront correction, several RL methods have been used successfully in conjunction with a pupil-plane WFS. These methods are currently the most efficient techniques among wavefront control approaches that leverage RL. To provide context for our study and highlight the broader landscape of RL in wavefront control, we begin this subsection with a review of the major pupil-plane WFS-based RL methods. 

Nousiainen et al. \cite{nousiainen_adaptive_2021, nousiainen_toward_2022, nousiainen_advances_2022, nousiainen_laboratory_2023} successfully used model-based RL to control large numbers of DM actuators at frequencies up to 1 kHz. Their closed-loop system was able to handle the non-linearities in the Pyramid WFS and adapt to changing conditions by self-calibrating. This approach surpassed the performance of classical integrator-based second-stage AO controllers in laboratory experiments. Although model-based methods are more sample-efficient (i.e. they require fewer interactions with the environment) than model-free methods, the latter have been shown to achieve comparable performance on a benchmark of continuous RL environments \cite{wang_benchmarking_2019}. 
Reference~\citeonline{landman_self-optimizing_2021} proposed a model-free method utilizing recurrent neural networks to capture temporal information from the Shack-Hartmann WFS measurements for improved turbulence prediction. To handle large action spaces, which are encountered when the number of DM actuators is high, they used convolutional layers in the agent architecture to identify local patterns and replaced the single global reward by a matrix of actuator-specific rewards. A similar method was applied to a Pyramid WFS, incorporating a combination of a linear and non-linear phase reconstruction as input to address the inherent non-linearities of the sensor. The non-linear phase reconstruction was obtained using an autoencoder network~\cite{pou_model-free_2022}.
To address the challenge of high-dimensional action spaces with model-free RL, \cite{pou_adaptive_2022} introduced a multi-agent approach, dividing the correction task into smaller sub-problems managed by individual agents. This approach, incorporating an autoencoder for noise reduction, achieves performance comparable to the state-of-the-art Linear Quadratic Gaussian controller, with the added benefit of being data-driven and capable of online training. Although they achieve state-of-the-art performance, these methods were designed for ground-based AO, and use pupil-plane WFS measurements as inputs, making them insensitive to some aberrations such as non-common path aberrations.

Concurrently, model-free RL has been investigated as a control strategy for WFS-less AO, i.e. AO systems that use focal plane images instead of a dedicated WFS for measuring wavefront distortions. Reference~\citeonline{ke_self-learning_2019} trained an agent to maximize image sharpness by analyzing features extracted from a single far-field intensity image using a CNN. The method achieves results comparable to a blind optimization (Stochastic Parallel Gradient Descent) and to a classical model-based WFS-less method~\cite{Lianghua:17}, but with a faster correction speed. However, this method requires 50 iterations to correct each wavefront distortion, which makes it too slow for realistic applications. 
A study in the field of biomedical optics~\cite{durech_wavefront_2021} demonstrated the effectiveness of RL in WFS-less AO for microscopy, by successfully correcting 5 low-order Zernike modes in more than 6 times fewer iterations than a traditional hill-climbing search (Zernike Mode Hill Climbing). Their method starts by combining the aberrated image with 10 additional images, which are generated by adding the maximum and minimum values of each individual mode to the initial aberrations. This initial observation is then followed by ``dynamic'' observations, containing the most recent actions and rewards. Although effective, this method is limited to the correction of a small number of low-order modes.
Finally, reference~\citeonline{parvizi_reinforcement_2023} trained an agent to maximize the Strehl ratio using a single focal plane image obtained with a low-pixel detector for WFS-less AO applied to Optical Satellite-to-Ground Communications. However, this method achieves a maximum Strehl ratio below 80\%, which isn't enough to compete with WFS-based approaches.
Moreover, although they eliminate the need for a WFS, the above methods \cite{ke_self-learning_2019, parvizi_reinforcement_2023} focus on correcting a limited number of low-order aberrations using a single intensity image, and thus do not explicitly lift the sign ambiguity in the aberrations. Both methods, requiring several tens of iterations for each correction, likely mitigate this ambiguity through a temporal approach, \emph{i.e.}, create the necessary diversity in their image sequence. Furthermore, the scalability of these methods to larger numbers of modes remains to be verified.

Unlike prior WFS-less RL approaches, we incorporate phase diversity information into the RL agent's observations. This strategy is advantageous because it is sensitive to all image-degrading aberrations and avoids the sign ambiguity issue. This allows for a more accurate correction compared to single-image techniques. In this work, we develop a model-free RL method for the correction of several tens of modes, and validate it on simulated realistic conditions.

\section{Theoretical background}
\label{sec:theory}

As this work is at the intersection of two communities (RL and High Angular Resolution imaging), we provide in this section a detailed introduction of the basics of both fields.

\subsection{Focal plane wavefront correction}

\begin{figure}[htbp!]
\centering\includegraphics[width=\myfraction\linewidth]{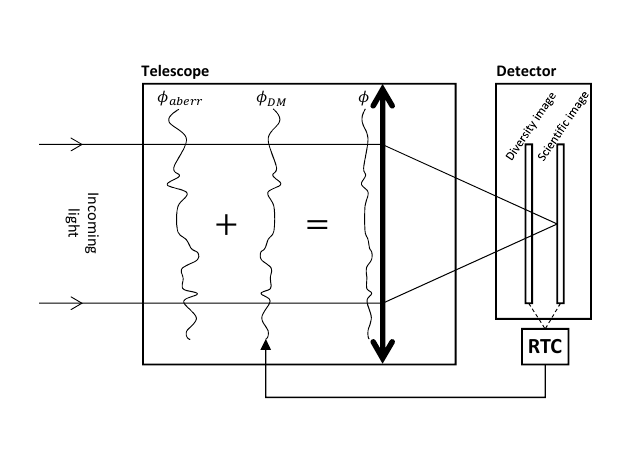}
\vspace*{-\baselineskip}
\caption{Principle of focal plane wavefront correction: phase diversity images are used to estimate wavefront aberrations. A real-time controller (RTC) then computes DM commands to correct these aberrations.}
\label{fig:wfc}
\end{figure}

For an optical telescope of aperture diameter $D$ observing a single astronomical source at wavelength $\lambda$, the theoretical angular resolution limit is $\lambda/D$. However, due to the presence of optical aberrations, this diffraction limit is in practice rarely reached. For large telescopes, a DM can be used to compensate for the wavefront aberrations and improve the imaging performance. This requires knowledge about the aberrations, which can be obtained through WFS measurements. In this section, we provide a quick introduction to focal plane wavefront sensing and correction, which traditionally uses a pair of phase diversity images to estimate the aberrations. A visual representation of the process can be found in Fig.~\ref{fig:wfc}.

We consider only one polarization state, and only phase aberrations. The source is assumed to be point-like at infinity, so that the incoming wavefront in the entrance pupil would be flat in the absence of aberrations. Under the Fraunhofer approximation, the intensity distribution in the imaging focal plane can be expressed as:
\begin{equation}
\label{eqn:psf}
    I\left(\Vec{x}\right) \propto \left|FT^{-1}\left[P\left(\Vec{\xi}\right)e^{i\phi\left(\Vec{\xi}\right)}\right]\right|^{2}\left(\Vec{x}\right)
\end{equation}
where $P$ is a function describing the transmission of the (scaled) pupil, which is non-zero only between $-D/ 2\lambda$ and $+D/ 2\lambda$, $\Vec{x}$ the angular position in the focal plane, $\Vec{\xi}$ the position in the pupil plane, $\phi$ the phase variation in the pupil plane at wavelength $\lambda$, and $FT^{-1}[\cdot]$ the inverse Fourier Transform operator.
The pupil plane phase is the sum of the entrance aberrations and the phase applied to the surface of the DM:
\begin{equation}
\label{eqn:phase}
\phi\left(\Vec{\xi}\right)= \phi_\text{aberr}\left(\Vec{\xi}\right) + \phi_\text{DM}\left(\Vec{\xi}\right)
\end{equation}
To completely correct the phase aberrations and thus reach the diffraction limit, the phase applied to the DM surface must therefore be the exact opposite of the phase aberrations.
The phase aberrations can be parameterized using Zernike polynomials. The Zernike polynomials form an orthonormal basis on the unit disk and thus offer a convenient mathematical framework for representing the phase distortions across the circular pupil of a telescope. Using this basis, the aberrations in the pupil plane at position $\Vec{\xi}$ can be written:
\begin{equation}
\label{eqn:aberrations}
\phi_\text{aberr}\left(\Vec{\xi}\right)= \sum_{i=2}^{\infty} c_{i}Z_{i}(\Vec{\xi}) 
\end{equation}
where $Z_i$ are the Noll-indexed Zernike modes \cite{Noll:76} and $c_i$ the corresponding coefficients. In practice, the phase is often expanded on a finite number of basis vectors, $N_\text{modes}$. The first mode, the piston, has no effect on the focal plane images and can thus be ignored. The following two terms, the tip and tilt modes, influence the position of the telescope's Point Spread Function (PSF). 

The DM can reproduce a finite number $N_\text{act}$ of Zernike modes thanks to its actuator array. In particular, the known aberration used for the diversity image (which is usually a defocus) can be obtained with the DM. The DM phase can be written as:
\begin{equation}
\phi_\text{DM}\left(\Vec{\xi}\right)= \sum_{i=2}^{N_\text{act}+1} d_{i}Z_{i}(\Vec{\xi})
\end{equation}
with $d_i$ the DM coefficients (not to be confused with actuator voltages).
The quality of the focal plane image depends on the residual root mean square (RMS) wavefront deviation
\begin{equation}
\sigma_{\phi}= \sqrt{\sum_{i=2}^{N_\text{modes}+1} (c_i + d_i)^2}
\end{equation}
where $d_{i, i>N_\text{act}+1}=0$.
To measure image quality, the Strehl ratio $\mathnormal{SR}$ is defined as the ratio of the peak intensity in the aberrated image over the maximum attainable intensity of a diffraction-limited optical instrument with the same aperture. This quantity can be directly obtained from the focal plane images. For low amplitude RMS wavefront distortions, the Strehl ratio can also be computed using the Mahajan approximation \cite{Mahajan:83}:
\begin{equation}
\label{eqn:Mahajan}
\mathnormal{SR} = e^{-\sigma_{\phi}^2}
\end{equation}
The maximum value $\mathnormal{SR}=1$ is obtained when $\sigma_{\phi}=0$, i.e. when all aberrations have been corrected. 

We mentioned in the introduction that phase retrieval from a single image suffers from a sign ambiguity in the estimation. This occurs for centro-symmetrical pupils, where a single image is not enough to retrieve the sign of the even part of the phase, as both solutions produce the same image in the focal plane. A detailed explanation of this phenomenon can be found, \emph{e.g.}, in Section~1.3 of \cite{Mugnier-l-06a}. 

Recording a second image with the same unknown phase along with an additional known even phase provides enough new information to lift the sign ambiguity and retrieve the complete phase. This so-called phase diversity approach frequently uses defocus as the additional diversity phase~\cite{Mugnier-l-06a}.

While it is possible to estimate aberrations using a single image under specific conditions, involving, \emph{e.g.}, an asymmetrical aperture~\cite{Baron-a-08,Martinache_2013} or a single defocused or otherwise diverse image~\cite{Meimon-a-10a}, we do not explore these scenarios in this study. Our work focuses solely on the context of phase diversity, where enough information is available to correct the aberrations with one measurement (\emph{i.e.}, one pair of images).

\subsection{Reinforcement Learning}

Reinforcement Learning \cite{sutton2018reinforcement} is a field of machine learning which consists in training an intelligent agent to make decisions in a dynamic environment in order to reach a goal. In this process, the agent receives rewards or penalties based on its actions, and tries to learn a policy that maximizes the cumulative sum of rewards received over time. The agent must balance between choosing the best action based on its current knowledge (exploitation), and trying out new options that may lead to better outcomes in the future (exploration), in order to learn an effective strategy in as few interactions with its environment as possible.

A Markov Decision Process (MDP) is a mathematical framework used to model RL problems. The decision-making entity being trained is called the \emph{agent}, and it interacts with an external system, containing everything outside the agent, called the \emph{environment}.  

An episodic MDP is an MDP where the agent-environment interaction can be broken down into finite sequences --- which we will refer to as \emph{episodes} --- of discrete \emph{time steps}. At each time step $t$, the agent receives a representation of the environment's current \emph{state} $s_{t} \in \mathcal{S}$, and uses this information to select an \emph{action} $a_{t} \in \mathcal{A}$ which is sent back to the environment. Upon receiving the action one time step later, the environment emits a \emph{reward} signal $r_{t+1} \in \mathcal{R} \subset \mathbb{R}$ and transitions to a new state $s_{t+1}$. The probability of transitioning from state $s \in \mathcal{S}$ to state $s' \in \mathcal{S}$ as a consequence of action $a \in \mathcal{A}$ is defined by the \emph{state-transition probability} function: 
\begin{equation}
\begin{aligned}
    p:\mathcal{S} \times \mathcal{S} \times \mathcal{A} &\to [0,1]\\
    s',s,a &\mapsto \Pr\{s_{t+1}=s' \,\mid\, s_{t}=s, a_{t}=a\}
\end{aligned}
\end{equation}
This probability must be completely independent of the past history of states and actions, for all $t'<t$. This is called the Markov property, and it facilitates the learning process in RL problems, as it allows the dynamics of the system to be deduced from the current state and action. An MDP is formally defined by the tuple $(\mathcal{S}, \mathcal{A}, p, \mathcal{R})$. 

In practice, the true state of the environment is not always fully accessible. The agent receives instead an \emph{observation} $o_{t} \in \mathcal{O}$. Observations provide indirect information about the true state, allowing the agent to make informed decisions despite not having full knowledge of the environment's state. 

In order to select the best action, the agent seeks to optimize a decision-making strategy or \emph{policy}, which is formally a mapping from states to a probability distribution over all possible actions:
\begin{equation}
\begin{aligned}
    \pi:\mathcal{S} \times \mathcal{A} &\to [0,1]\\
    s,a &\mapsto \Pr\{a_{t}=a \,\mid\, s_{t}=s\}
\end{aligned}
\end{equation}
The policy outputs the probability of taking action $a$ when in state $s$. The algorithm chosen to optimize the policy will be described in \Cref{subsec:ppo}.

\section{Method}
\label{sec:method}

This section gives an overview of the method that we developed. In Subsection \ref{subsec:formulation}, we leverage the concepts introduced in \Cref{sec:theory} to formulate the focal plane wavefront correction problem within the framework of RL. In Subsection \ref{subsec:ppo}, we explain the reasoning behind our choice of optimization algorithm, and Subsection \ref{subsec:hyperparameters} details the implementation process and the rationale behind our choice of training hyperparameters.

\subsection{Focal plane wavefront control as a Reinforcement Learning problem}
\label{subsec:formulation}
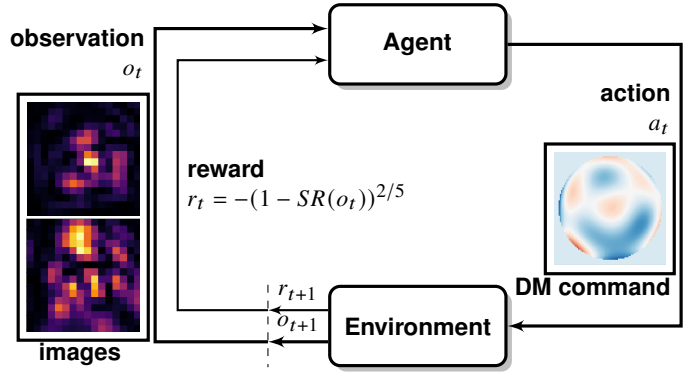
\begin{figure}[htbp!]
\centering
\begin{tikzpicture}[font=\small\sffamily\bfseries,very thick,node distance = 4cm, every label/.style={align=center}]
\node [frame] (agent) {Agent};
\node [frame, below=2.65cm of agent] (environment) {Environment};
\draw[line] (agent) -- ++ (3.5,0) |- (environment)  node[left, pos=0.115, align=right] {action\\ $a_t$}
node[mynode,pos=0.41, anchor=south east] {\includegraphics[scale=0.16]{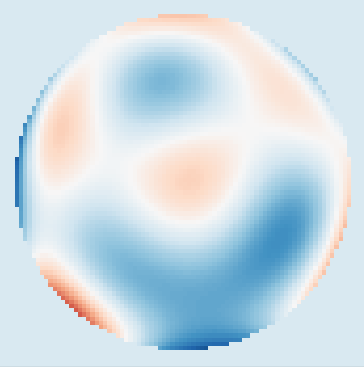}} node[left, pos=0.43] {DM command};
\coordinate[left=8mm of environment] (P);
\draw[thin,dashed] (P|-environment.north) -- (P|-environment.south);
\draw[line] (environment.190) -- (P |- environment.190)
node[midway,above]{$o_{t+1}$};
\draw[line,thick] (environment.170) -- (P |- environment.170)
node[midway,above]{$r_{t+1}$};
\draw[line] (P |- environment.190) -- ++ (-1.5,0) |- (agent.170) node[left, pos=0.46, align=right] {observation \\ $o_t$}
 node[mynode, pos=0, anchor=south east] {\includegraphics[scale=0.18]{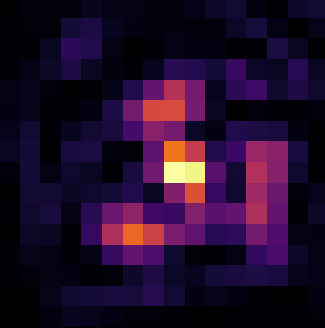} \\ \includegraphics[scale=0.18]{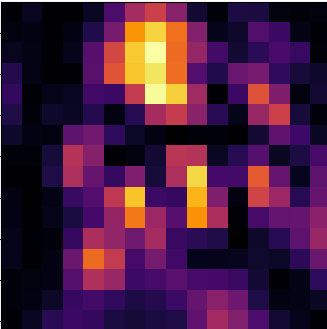}} node[left, pos=-0.02] {images$\,\,\,\,\,\,$};
\draw[line,thick] (P |- environment.170) -- ++ (-1.2,0) |- (agent.190)
node[right,pos=0.25,align=left] {reward\\ $r_t = -(1 - \mathnormal{SR}(o_{t}))^{2/5}$};
\end{tikzpicture}
\caption{Diagram of the interaction between the agent and its environment. At each time step $t$, the environment emits the observation $o_t$, containing the in-focus (top) and out-of-focus (bottom) images, as well as the reward $r_t$. The agent responds with action $a_t$ (the DM command) which prompts the environment to emit a new observation and reward $o_{t+1}$ and $r_{t+1}$, and so on.}
\label{fig:rlwfc}
\end{figure}

Our formulation of the focal plane wavefront correction problem as an episodic MDP is illustrated in Fig.~\ref{fig:rlwfc}. In the case of focal plane wavefront control, the agent is the DM controller. The environment is everything outside the controller, including the instrument and its dynamics, as well as the incoming distorted wavefront. The state $s_{t}$ corresponds to the aberrations in the pupil plane ($\phi$ in Eq. (\ref{eqn:phase})), which cannot be measured directly. Instead, the agent observes two focal plane phase diversity images, which contain all the information necessary to estimate the aberrations. The agent uses this observation  $o_{t}$ to select an action $a_{t}$, which here is the DM phase $\phi_\text{DM}$. The reward $r_{t+1}$ is computed from the next observation, and gives a measure of the quality of the correction. We chose a negative increasing function of the Strehl ratio $\mathnormal{SR}(o_{t+1})$ in the in-focus image (here, the reward is independent of the action taken):
\begin{equation}
\begin{aligned}
    r:\mathcal{O} &\to \mathcal{R}=[-1,0]\\
    o &\mapsto -(1 - \mathnormal{SR}(o))^{2/5} 
\end{aligned} 
\end{equation}

The design of the reward function allows for some flexibility. It can be chosen freely, as long as it can be computed at each time step and the maximization of the reward is equivalent to correcting all aberrations; nevertheless, not all functions meeting this requirement are equally effective. Here the reward function is maximal when the aberrations are minimal, and the Strehl ratio can be computed at each time step by using the theoretical PSF of the instrument. The choice of a negative reward urges the agent to try and find the best correction as quickly as possible, since each time step where a non-zero reward is received acts as a penalty, lowering the overall score the agent is trying to accumulate. The shape of the function has been chosen to encourage the agent to explore a wide variety of actions when the Strehl ratio is low, and to choose greedier actions (i.e. actions that it predicts to yield the highest reward based on its current knowledge) as it gets close to 1. Designing reward signals is a cutting-edge research area, and this may not be the best reward function for this particular problem. The intuition for this choice came from Section 17.4 of \cite{sutton2018reinforcement}; also, of the several reward functions tested experimentally, this one led to the highest performance in practice. 

The importance given to future rewards is determined by the \emph{discount rate} $0\leq\gamma\leq1$. If $\gamma=0$, the agent aims to maximize the immediate reward; as $\gamma$ approaches 1, the future rewards are taken into account more strongly. In this setup, after experimenting with several values of $\gamma$, we ultimately found that $\gamma = 0$ led to the best training performance. This is consistent with the expected behavior of phase diversity, since each observation provides enough information to completely correct the aberrations with a single DM command, meaning we are solely interested in the Strehl ratio immediately following each action. Note that if the observations contained a single focal plane image,  then $\gamma$ should probably be closer to 1 as the agent would need to learn to first create its own diversity image with a sub-optimal command, sacrificing immediate reward to get a better estimate of the aberrations.

The choice of episode length depends on several factors, which will be discussed more in-depth in \Cref{subsec:epsizes}. 

\subsection{The Proximal Policy Optimization algorithm}
\label{subsec:ppo}

The algorithm used to train the agent belongs to the Proximal Policy Optimization (PPO) algorithms \cite{SchulmanWDRK17}. PPO is a variant of \emph{actor-critic} algorithms \cite{konda1999actor}, which belong to the \emph{policy gradient} methods family. These methods learn a parameterized policy \cite{sutton_policy_1999} that can select actions without using a value function.  Actor-critic methods learn both the policy and the state-value function in parallel, the ``actor'' referring to the learned policy, and the ``critic'' to the learned state-value function. The critic provides a feedback to the actor which helps improve the decision-making over time \cite{mysore2021honey}. 

In practice, PPO algorithms alternate between filling a \emph{rollout buffer} with data generated through interaction with the environment, and optimizing the policy on this rollout buffer using stochastic gradient ascent. Additionally, PPO constrains the policy updates to a specified range around the old policy. This range acts as a form of trust region, preventing large policy changes that could lead to instability.
The motivation for this choice is twofold: on the one hand, PPO benefits from the stability and reliability of trust region methods such as TRPO \cite{schulman2015trust}, while being simpler to implement, as well as having better empirical generalization and sample complexity \cite{SchulmanWDRK17}. On the other hand, PPO has shown good performance on a variety of tasks where the action space is continuous \cite{li2023deep}, which is the case here. However, although we have found PPO to be well-suited to the task at hand in this paper, we do not claim that PPO is the best for this particular problem. In particular, because the action yielding the highest immediate reward is always the optimal choice (see \Cref{subsec:formulation}), the critic might have limited utility in this specific scenario; simpler policy-gradient algorithms could potentially be as effective for this problem. A more extensive study should be conducted to determine the most efficient training method.

\subsection{Training configuration}
\label{subsec:hyperparameters}

We trained the agent using a Python implementation of PPO provided by the \texttt{Stable-baselines3} library \cite{stable-baselines3}. 
We use the Adam optimizer \cite{kingma2017adam} with a starting learning rate of $10^{-3}$ and a scheduler linearly decreasing the global learning rate towards $10^{-3}/T_{total}$, with $T_{total}$ the total number of training time steps. This approach asymptotically approximates a $1/t$ decay while providing a gentler slope in the earlier stages of training to promote exploration. The global learning rate serves as an upper limit to Adam's adaptive learning rate, and we found that adding a linear decay improved the stability of the training, and led to faster convergence in practice. The other hyperparameters for Adam are left at their default values. 

For the PPO hyperparameters, we chose a rollout buffer size of 8000 time steps, and a mini-batch size of 500 time steps. Other parameters (clipping parameter, epochs, Generalized Advantage Estimator decay parameter, advantage normalization, value function coefficient and entropy coefficient) can also be modified, but we found that their default value from \texttt{Stable-baselines3} yielded the best results in our experiments. 

\begin{figure}[htbp!]
\centering\includegraphics[width=\myfraction\linewidth]{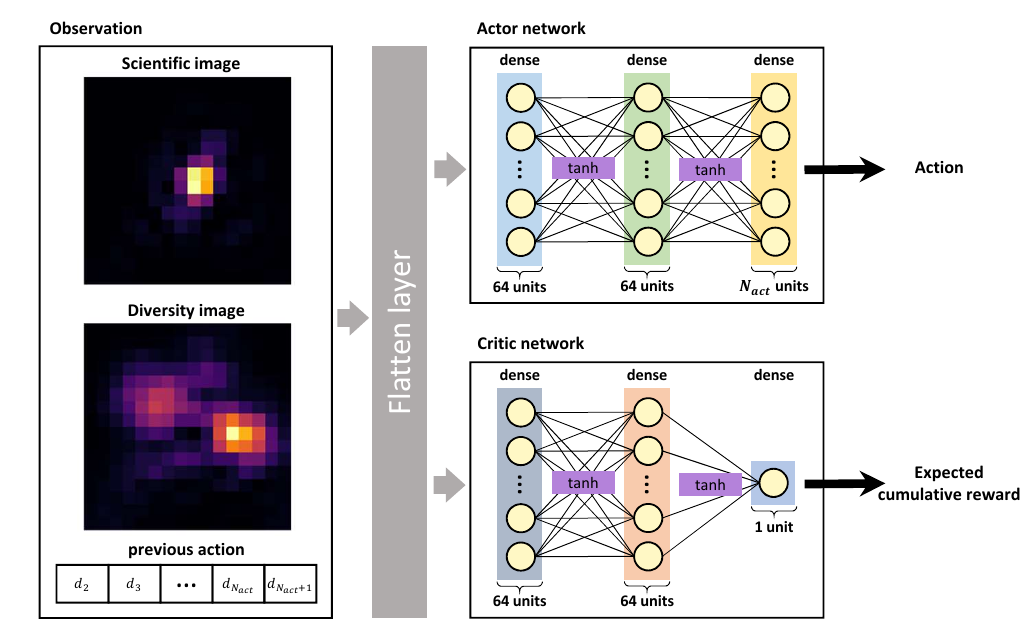}
\caption{Actor and critic network architectures.}
\label{fig:architectures}
\end{figure}

Both the actor and critic are implemented as neural networks. The specific architectures of these networks are represented in Fig.~\ref{fig:architectures}. The actor and the critic share the same architecture, with different output layers. The input is the observation $o_t$, where both images are flattened into a single 1-D vector, to which we append the previous DM command $[d_{2,t-1}, ..., d_{N_\text{act}+1,t-1}]$ (with $d_{i,t=0} = 0, \forall i$). The data is processed by two sequential fully connected hidden layers, each containing 64 neurons. The activation function of the hidden layers is the hyperbolic tangent (tanh) function. Finally, the output layer of the actor contains $N_\text{act}$ neurons, corresponding to the number of DM actuators, while the critic network employs a single output neuron dedicated to estimating the expected cumulative reward from the current state. The output command is a vector of coefficients for the Zernike corrector $[d_{2,t}, ... d_{N_\text{act}+1,t}]$, with $d_{i,t} \in [-1,1]$, $\forall i$. The $d_i$ are then scaled by $\underset{2\leq i\leq N_\text{modes}+1}{\max} \max(c_i)$, with $c_i$ the Zernike coefficients as in Eq.~(\ref{eqn:aberrations}), to ensure the DM can always correct the full range of aberration amplitudes. In practice, since the $c_i$ are unknown, the $d_i$ would be scaled by the maximum expected value for each mode or the maximum stroke of the DM actuators. Within our simulated environment, however, we have access to the maximum amplitude across all aberration coefficients, which we use as a prior for the DM coefficients in order to focus solely on potential issues arising from the learning process itself.

Further performance improvements might be achievable through hyperparameter and architecture optimization, but this exploration is beyond the scope of this work.

\section{Numerical simulations}
\label{sec:simulations}

To evaluate the performance of our method under realistic conditions, we designed a reference experiment, termed ``control case'' in the following, which we describe in this subsection.

The environment is defined using OpenAI's Gym package \cite{brockman2016openai}, and simulations are carried out using the Asterix package in Python \cite{mazoyer_AsterixSimulator}. 

\begin{figure}[htbp!]
\centering\includegraphics[width=\myfraction\linewidth]{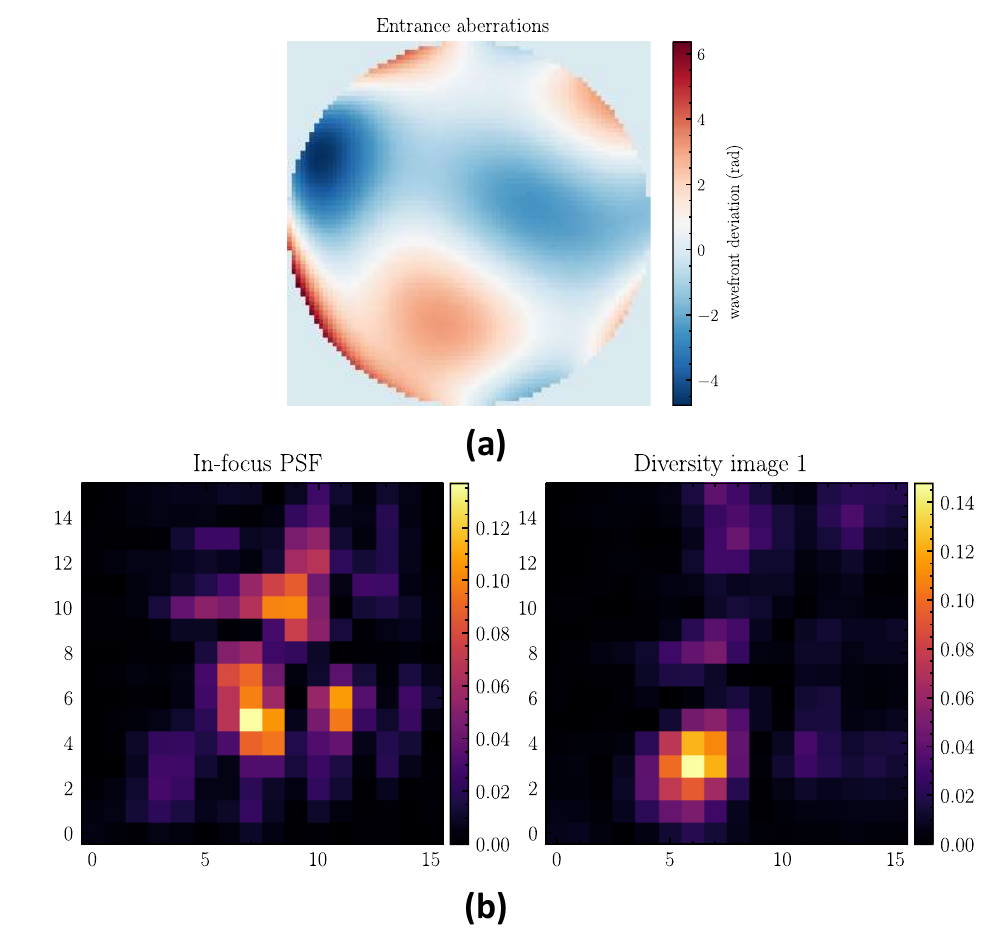}
\vspace*{-.5\baselineskip}
\caption{Example of an entrance phase map (a) and the resulting in-focus (left) and out-of-focus (right) images observed by the agent (b).}
\label{fig:examples}
\end{figure}

Observations are conducted at a wavelength of $\lambda=500\,\text{nm}$. For the control case, we aim to correct $N_\text{modes}=21$ simultaneously, i.e. all modes (except the piston) up to and including the secondary spherical aberration, achieving a full odd radial order as well as including the isotropic mode from the next even order (the modes are numbered using the Noll indexing \cite{Noll:76}). The Zernike coefficients are uniformly sampled within the range $[-\max(c_i),\max(c_i)]$, where $\max(c_i)$ is computed for each mode of index $i$ such that the aberrations follow a $1/f^2$ ($f$ denoting the spatial frequency) spatial Power Spectral Density, which has been measured for high-quality optical surfaces \cite{dohlen2011}. The coefficients are then scaled so as to result in a total RMS wavefront deviation over the aperture of $125\,\text{nm}$ (i.e. $\lambda/4$) before correction. Additionally, we capped the tip and tilt errors to $+/- 0.5$ $\lambda/D$ around the center to emulate an upstream geometric beam centering with an accuracy of around 1 pixel.
The entrance aberrations are considered to be static for the duration of an episode, and the agent is presented with a new independent phase screen at the start of each episode. An example of phase map produced with these settings is shown in Fig.~\ref{fig:examples} (a).

The telescope itself has a circular, unobstructed pupil. The photons are registered on the $16\times16$-pixel grid of a square detector, such as a CCD camera. The pixel pitch is adjusted so that two pixels make up one $\lambda/D$, which results in images sampled at the limit given by the Nyquist-Shannon sampling theorem. The detector field of view is thus $8\times8$  $\lambda/D$. 

A defocus with a peak-to-valley amplitude of $\lambda$ (corresponding to $\lambda/(2\sqrt{3})$ RMS) is introduced in the pupil plane to obtain the out-of-focus image. All images are normalized so that the maximum intensity would be one in the absence of aberrations.
Fig.~\ref{fig:examples} (b) shows the diversity images obtained from the phase map in Fig.~\ref{fig:examples} (a).

The correction is performed by a single DM. The DM is an ideal Zernike corrector with $N_\text{act} = N_\text{modes}$ degrees of freedom. This simulated device can perfectly reproduce any combination of the $N_\text{modes}$ Zernike modes comprising the entrance aberrations, i.e. it can completely cancel out the entrance aberrations provided the correct command is issued. The response time of the DM is instantaneous.
This setup allows to easily determine where the errors come from if the training goes wrong: since the DM can fully reconstruct the phase, which is fully observable by the agent (albeit indirectly and assuming no noise) thanks to the diversity image, any correction that doesn't yield a final Strehl ratio of $100\%$ is the result of an error in either the estimation of the environment state or the action that was chosen. Either way the agent is at fault and the error should be fixed by using more training data or fine tuning the training hyperparameters (see \Cref{subsec:hyperparameters} for the chosen hyperparameters).

To evaluate our method's noise tolerance (\Cref{subsec:noise}, we also introduce noise sources in the simulations. We model the detection system as photon noise limited, additionally including a Gaussian detector readout noise with a standard deviation $\sigma_\text{RON}$ of one electron. All other potential noise sources are considered negligible in this context. The measured intensity $I_\text{mes}$ in a pixel at coordinates $(x,y)$ under the influence of noise, before image normalization, can thus be expressed as:
\begin{equation}
I_\text{mes}(x,y) = \mathcal{P}\left[N_\text{ph}(x,y)\right] + n_\text{RON}(x,y)
\end{equation}
with $N_\text{ph}$ the number of photoelectrons collected in the pixel before application of the noise, and $\mathcal{P}\left[N_\text{ph}\right]$ denotes the Poisson distribution of the photon noise, and $n_\text{RON}$ the detector readout noise, following a 0-mean i.i.d. Gaussian distribution of variance $\sigma_\text{RON}^2$. $N_\text{ph}(x,y)$ is linked to the signal-to-noise ratio in the pixel $SNR_\text{pix}=N_\text{ph}(x,y)/\sqrt{N_\text{ph}(x,y) + \sigma_\text{RON}(x,y)^2}$. We define the``global signal-to-noise rati'' of the experiment as the value of $SNR_\text{pix}(0,0)$ in the un-aberrated PSF (i.e. the value in the brightest pixel in the absence of aberrations). Therefore, the latter value is an upper limit for $SNR_\text{pix}$ across all pixels, and from now on we will refer to it simply as ``SNR'' for brevity. The SNR is set to 100 for the reference experiment.

Finally, the reference episode length is set to 4 time steps per episode. The reason behind this choice is discussed in \Cref{subsec:epsizes}. 

In the next Section, we evaluate the impact of varying key experimental parameters (episode length, number of modes and noise) on the learning process. 

\section{Results and discussion}
\label{sec:results}

In this section, we showcase and analyze the performance of our method. The results presented here were obtained by training the agent over 10 million time steps (which corresponds to 2.5 million training episodes for the reference experiment). Training was carried out on a 56-core CPU with a clock speed of 2.4 GHz and an Nvidia Tesla P100-PCIE-12GB GPU. The average training time was around 10 hours. 
Subsection \ref{subsec:control} contains the learning curve for the reference experiment. We study the influence of varying key parameters in subsections \ref{subsec:epsizes}, \ref{subsec:nmodes},  \ref{subsec:amplitudes} and \ref{subsec:noise}. Finally, we study the learning curve obtained when the aberrations are of higher order than what the DM can compensate for in subsection \ref{subsec:other}.

\subsection{Reference experiment}
\label{subsec:control}

\begin{figure}[htbp!]
\centering\includegraphics[width=\myfraction\linewidth]{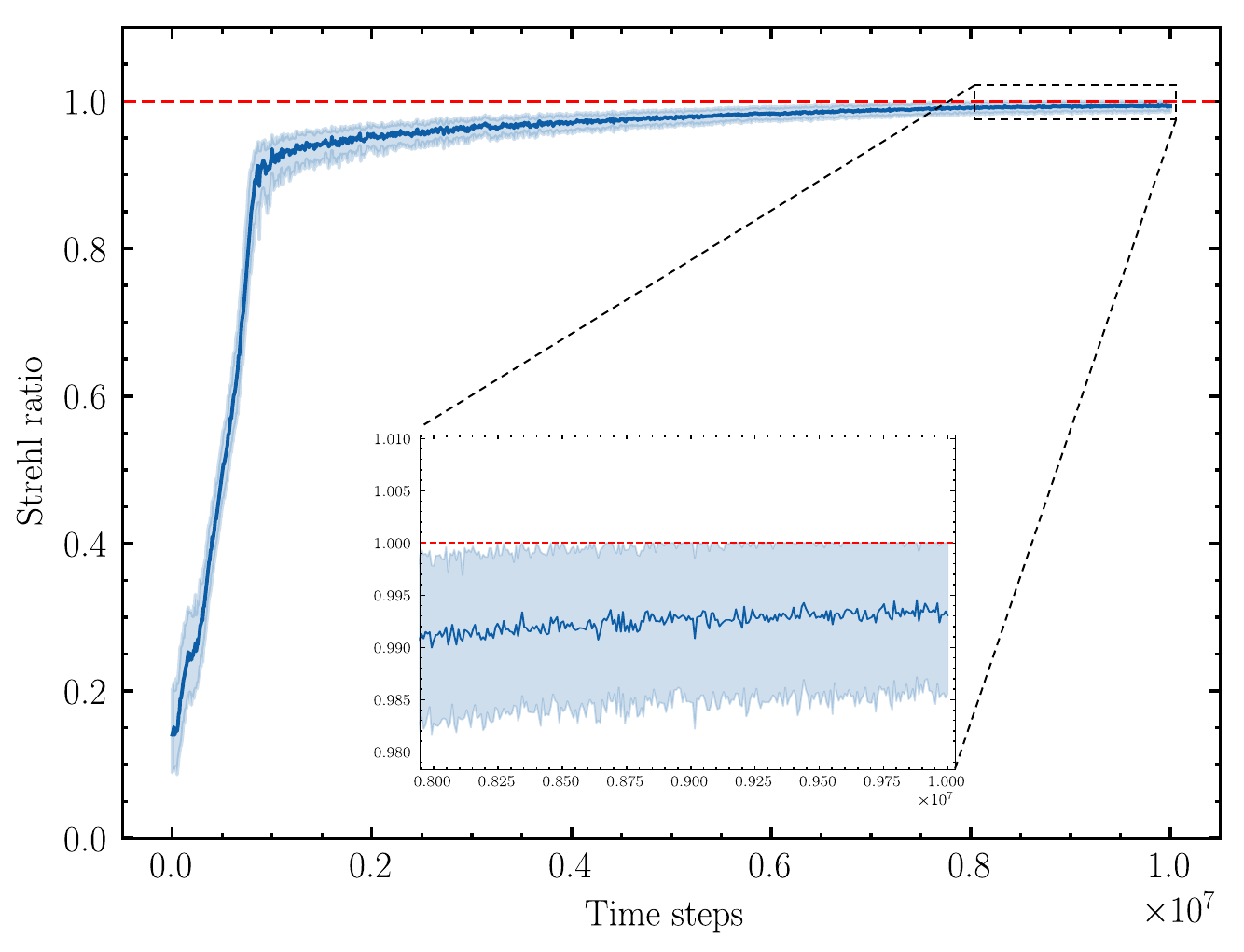}
\vspace*{-.5\baselineskip}
\caption{Average Strehl ratio after correction on the benchmark as a function of the number of training time steps, for the control experiment. The error bands show the spread of the results between the 25th and 75th percentiles. The agent's performance gets progressively closer to the optimal theoretical performance (represented by the red dashed line), reaching an average Strehl ratio greater than 0.99 after 10 million time steps.}
\label{fig:learningcurve}
\end{figure}

The evolution of the agent's performance over the course of training for the reference experiment is presented in Fig.~\ref{fig:learningcurve}. The performance is assessed after each gradient optimization step by testing the agent on a set of 100 entrance phase screens, running the policy for a full episode on each, and averaging the end-of-episode Strehl ratio across all phase screens. This set of phase screens remains constant throughout the entire training process, serving as a benchmark that enables accurate tracking of the agent's progress. The learning curve is averaged over 5 training runs to estimate the variance of the learning process, with the error bands showing the spread of the data between the 25th and 75th percentiles over the 5 runs. The results show that the agent is able to consistently obtain an average Strehl ratio of 0.9 on the benchmark after 2 million time steps, after which the progress slows down due to the linear decay of the learning rate, the Strehl ratio eventually exceeding 0.99 near the end of training. If we remove this linear decay, the learning process becomes unstable after the first 2 million time steps, 
and often results in the sudden performance collapse known as \emph{catastrophic forgetting}.

This result shows that the agent is able to autonomously learn to estimate and correct 21 pupil plane aberrations in realistic conditions (SNR=100, RMS wavefront deviation of $\lambda/4$) using only the available observations, and without relying on any prior knowledge about the aberrations. 

\subsection{Episode length}
\label{subsec:epsizes}

\begin{figure}[htbp!]
\centering\includegraphics[width=\myfraction\linewidth]{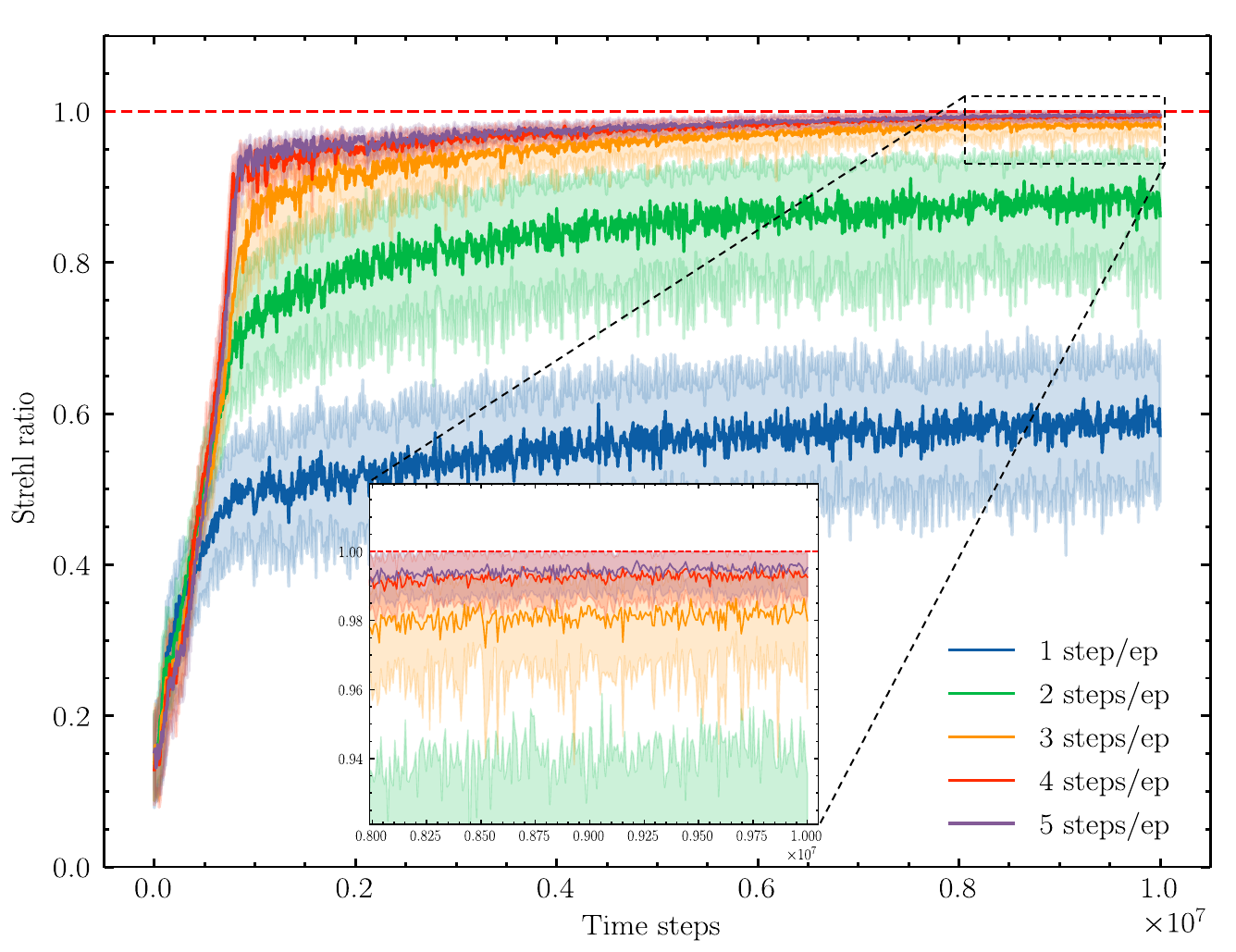}
\vspace*{-.5\baselineskip}
\caption{Learning curves of the agent when varying the episode length while keeping all other parameters at their nominal value. Each curve corresponds to a training performed with a specific episode length, ranging from 1 to 5 time steps per episode. The error bands show the spread of the results on the benchmark between the 25th and 75th percentile.}
\label{fig:epsizes}
\end{figure}

Episode length (i.e. the number of time steps per episode) is an important factor in the learning process. The ideal episode length is 1: if the agent is able to correct the aberrations with only one action, there is no need to acquire an additional pair of phase diversity images, and therefore no risk of the entrance aberrations evolving before the correction is finished. However, we found that the greater the episode length, the faster and more stable the learning process, and when the episode length is too low, the agent doesn't learn at all. The value of 4 time steps per episode was chosen for the control case because it is the lowest one that allows a stable training. Fig.~\ref{fig:epsizes} shows the influence of varying the episode length with all other parameters fixed at their nominal values. Reducing the episode length below a threshold of 4 time steps per episode leads to increased instability during training and limits the overall performance. On the flip side, increasing the number of time steps per episode only leads to marginal improvements. 

Our proposed interpretation is because the agent encounters phase variations larger than $2\pi$ (which corresponds to a wavefront deviation of $\lambda$), for which the phase diversity images do not give complete information. Indeed, because the phase appears in a complex exponential in Eq.~(\ref{eqn:psf}), a variation of $2\pi$ at a given point in the pupil plane phase has no effect on the focal plane images. Therefore, the agent's observations only contain information on the wrapped phase $\phi\left[2\pi\right]$. The nominal value of the RMS wavefront deviation ($125\,\text{nm}$, i.e. $\lambda/4$) considered in this study results in an average peak-to-valley amplitude of the entrance phase maps exceeding $2\pi$. Without any additional information, this creates an ambiguity in the agent's state representation, because it is harder (though not impossible) to discriminate between the actual continuous phase and the other equivalent phases. Allowing the agent to take several actions lets it learn this ambiguity by trying out possible solutions. To verify this hypothesis, we trained an agent on small amplitude aberrations, where the peak to valley amplitude cannot exceed $2\pi$. This corresponds to an RMS wavefront deviation of $50\,\text{nm}$, or $\lambda/10$. In this case, the agent was able to learn an effective control strategy with an episode length of 1. The experiment was repeated by training agents with episode lengths between 1 and 10 steps, while slowly increasing the RMS deviation such that each increment increases the maximum peak-to-valley amplitude by $2\pi$. The results are shown in \Cref{tab:pv_epsize}. Up to $167\,\text{nm}$ RMS (i.e. $\lambda/3$), the minimum episode length required for the agent to learn an effective strategy can be approximated as the maximum number of ``wrappings'' of the phase, i.e. as the quotient of the division of the maximum peak-to-valley amplitude of the training phase screens by $2\pi$, plus one. However, we do not observe this behavior at larger aberration amplitudes. In such cases, increasing the episode length still results in better average performance, but the chosen hyperparameters do not seem adequate anymore and the learning process is unstable. This behavior can be attributed to a non-linear growth in problem complexity as aberration amplitudes increase, though understanding this limitation falls outside the scope of this study. 

Another option that could enable proper learning of the agent would be to incorporate prior knowledge about the continuous nature of the true aberrations in the shape of a regularization term, but we leave that to further research.

An alternative explanation for why the agent learns better when increasing the episode length is that it allows the agent to explore the impact of its actions locally around the same initial state. This could be interpreted as the agent learning the local shape of its performance criterion, i.e. the PPO loss in this case. As remarked by one of the reviewers, the increased episode length also exposes the agent to a wider range of aberration amplitudes as its actions influence the RMS value of the residual phase in the pupil plane. However, while longer episodes offer benefits, they come at the cost of encountering fewer unique aberration shapes. Since the total number of time steps remains constant across experiments, increasing the number of time steps per episode reduces the number of individual episodes (and hence, the number of unique aberrations) experienced by the agent. This trade-off between diversity in aberration shapes and diversity in aberration amplitudes warrants further investigation.

\begin{table}[htbp!]
\centering\includegraphics[width=\myfraction\linewidth]{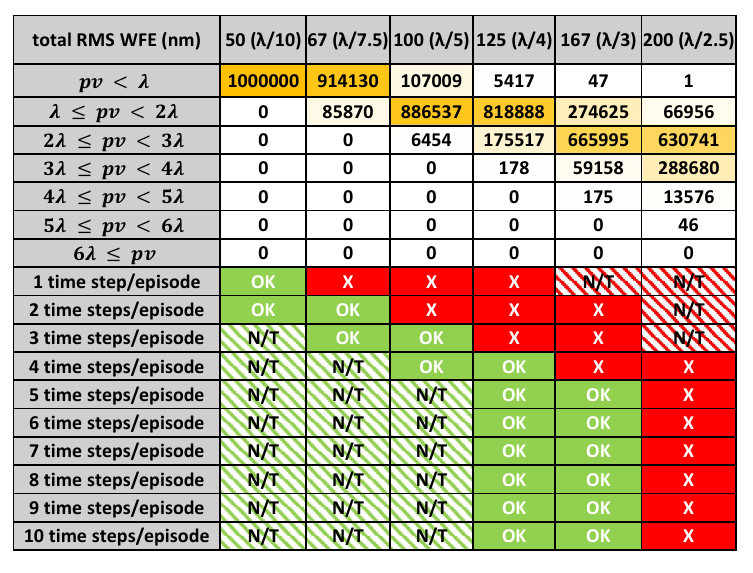}
\caption{Influence of the RMS wavefront error (WFE) on the amount of time steps per episode required to learn an effective control strategy. For a given RMS WFE, we generate 1 million phase screens following the distribution described in \Cref{sec:simulations} and count the number of peak-to-valley amplitudes (pv) in each ``wrapping''. The learning curves are summarized for each episode length with ``OK'' if the average Strehl ratio obtained at the end of training exceeds $0.95$, and ``X'' otherwise.  Green (respectively red) hashed cells show untested but likely successful (respectively unsuccessful) cases, with the mention ``N/T'' standing for ``Not Tested''.} 
\label{tab:pv_epsize}
\end{table}  

\subsection{Scaling to higher order aberrations}
\label{subsec:nmodes}

\begin{figure}[htbp!]
\centering\includegraphics[width=\myfraction\linewidth]{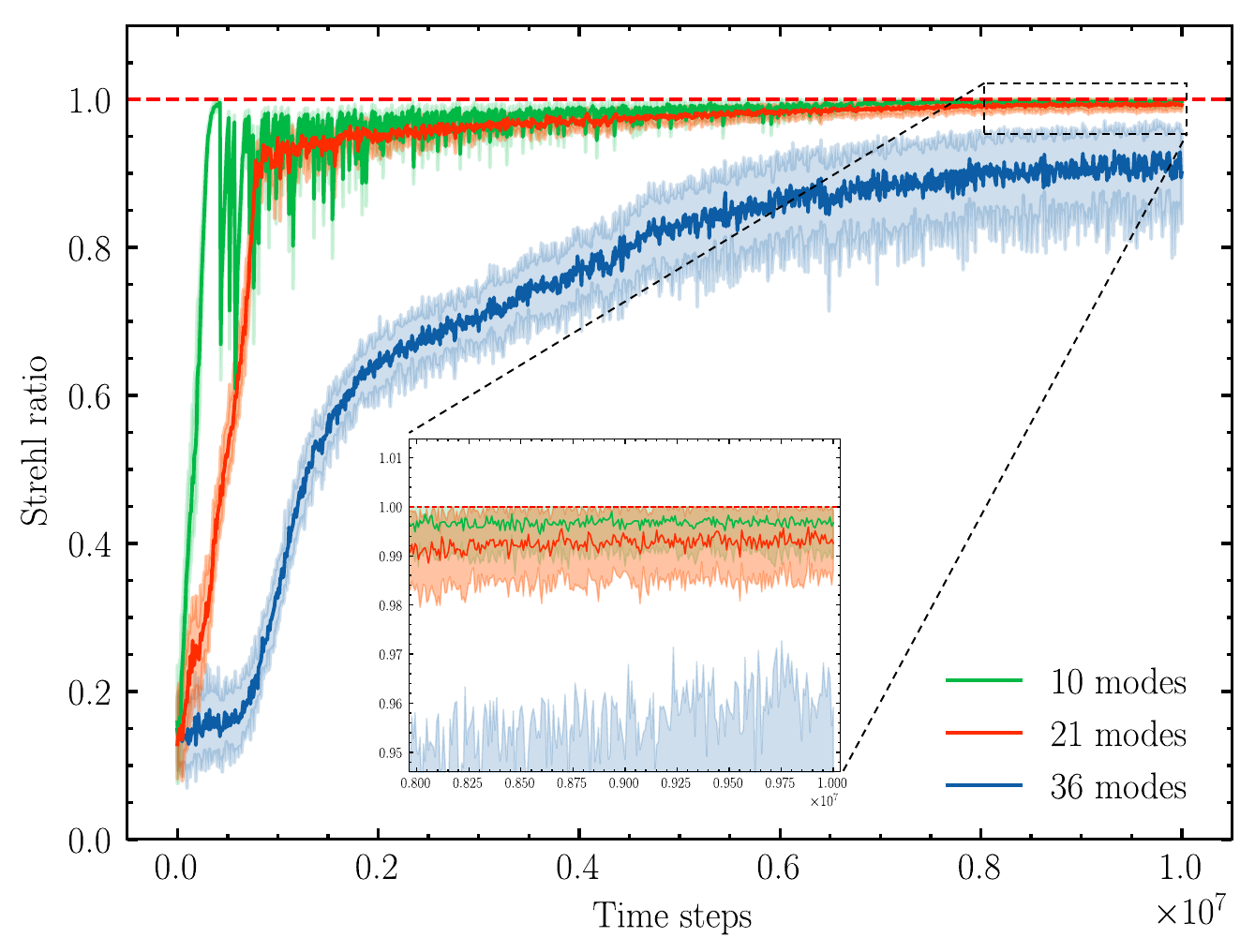}
\vspace*{-.5\baselineskip}
\caption{Learning curves for $N_\text{modes}=10$, $21$ and $36$, with all other parameters at their nominal value. In particular, the episode length has been set to 4 steps/episode. The error bands show the spread of the results on the benchmark between the 25th and 75th percentile.}
\label{fig:nmodes}
\end{figure}

In this section, we measure the influence of varying the number of modes on training and performance. Fig.~\ref{fig:nmodes} compares the control case learning curve with the learning curves for $N_\text{modes}=10$ and $36$ (i.e. all modes (excluding the piston) up to and including, resp. the primary spherical aberration and the tertiary spherical aberration), with all other parameters fixed at their nominal values. With $N_\text{modes}=10$, the agent is able to learn quickly and the average Strehl ratio after correction exceeds 0.99 long before the end of training. The fluctuations for $10$ modes at the early stages of training are likely due to the global learning rate still being too high when the agent approaches the maximum correction performance on the benchmark for the first time. Once the linear decay mechanism has sufficiently reduced the learning rate, performance stabilizes. For $N_\text{modes}=36$, training is too slow to reach an acceptable performance after 10 million training time steps. This is likely due to the increase in the number of degrees of freedom of the problem, which relates to an increase in the dimensions of both the state space and the action space. This results in a larger policy space to explore, which could explain the slower training. Likewise, reducing the number of modes to correct results in a faster training, as the agent requires less exploration to find a good policy. 

\begin{figure}[htbp!]
\centering\includegraphics[width=\myfraction\linewidth]{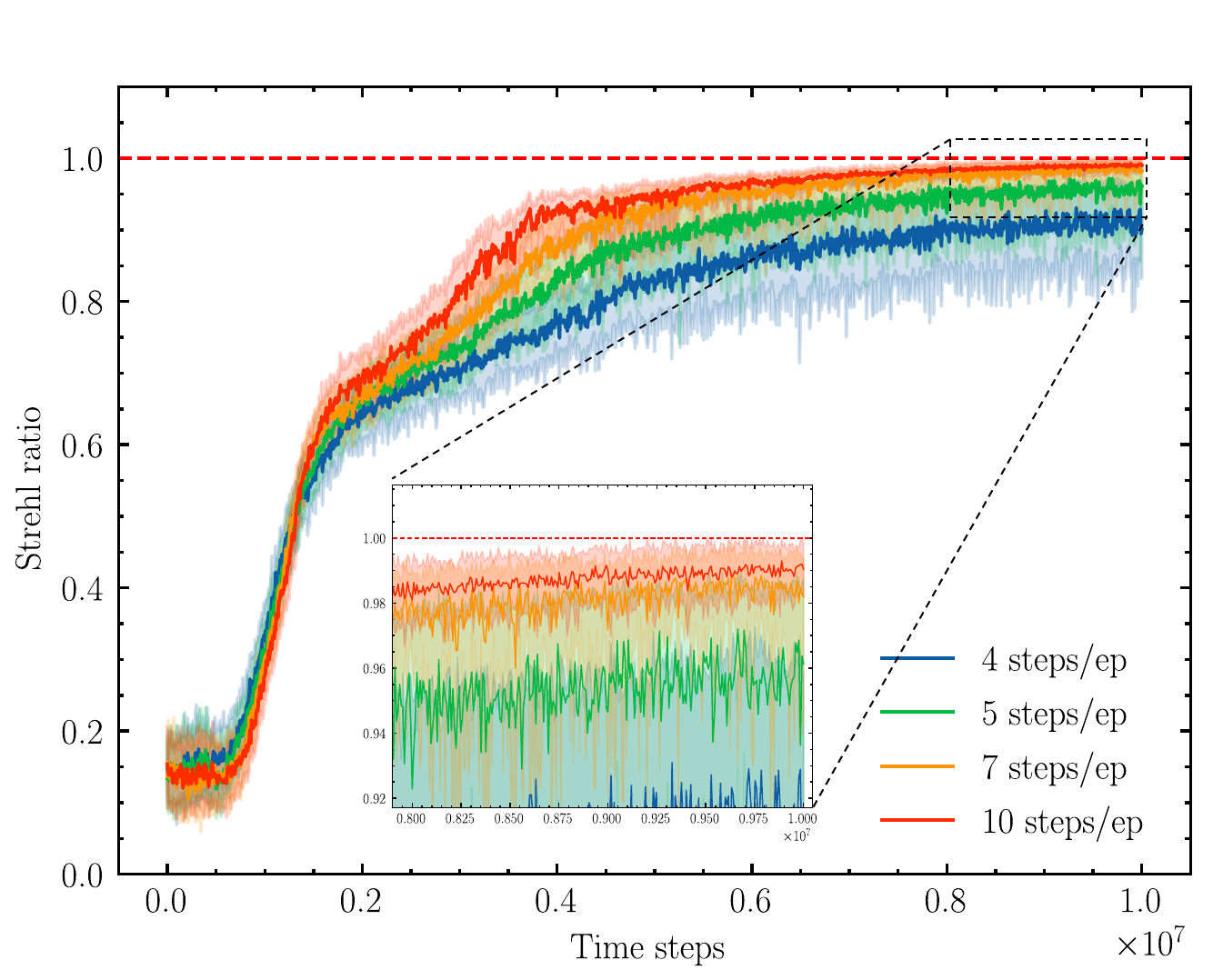}
\vspace*{-.5\baselineskip}
\caption{Learning curves for episode lengths ranging from 4 to 10 steps/episode, with $N_\text{modes}=36$, the error bands displaying the spread of the results on the benchmark between the 25th and 75th percentile. Increasing the number of time steps per episode accelerates the learning process and allows the agent to reach higher Strehl ratios by the end of training.}
\label{fig:36modes}
\end{figure}

Increasing the episode length allows the agent to perform better for higher numbers of modes (Fig.~\ref{fig:36modes}). We hypothesize that by incorporating more time steps per episode, the agent may learn to decompose the task into a sequence of incremental corrections to ease the learning process. Increasing the number of time steps per episode from 4 to 10 allows the method to work with up to 36 degrees of freedom. 

As already mentioned, using long episodes is not ideal because of the temporally evolving nature of the aberrations. As a workaround, one might benefit from allowing a greater episode length during the earlier stages of training, and gradually shortening it towards the end to combine the advantages of both alternatives.

\subsection{Generalization to higher or lower aberration levels}
\label{subsec:amplitudes}

This subsection explores the agent's ability to generalize beyond the specific aberration level it trained on. We investigate how well an agent trained on a fixed RMS aberration amplitude performs when encountering different aberration levels.

\begin{figure}[htbp!]
\centering\includegraphics[width=\myfraction\linewidth]{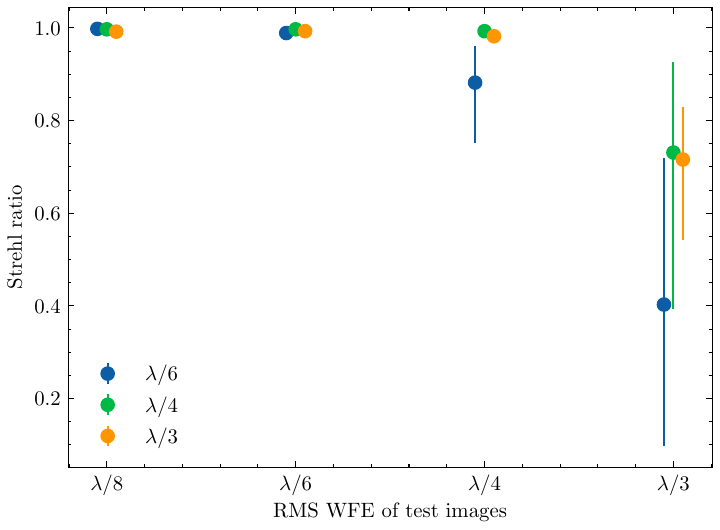}
\vspace*{-.5\baselineskip}
\caption{Average end-of-episode Strehl ratio obtained by the 3 agents as a function of the tested RMS values. The error bars represent the spread of the results between the 25th and 75th percentile. Error bars are smaller than the point width for $\lambda/8$ and $\lambda/6$ test cases.}
\label{fig:rms}
\end{figure}

To this end, we train an agent using the the same parameters as the reference experiment, with the sole modification being the re-scaling of the DM coefficients $d_i$ (see \Cref{subsec:hyperparameters}) so that they can cover a wider range of amplitudes, up to $\lambda/3$ RMS. This re-scaling does not impact training on the reference experiment, as the agent still achieves an average end-of-episode Strehl ratio of 0.99 on the benchmark at the end. 
Following training, the agent is presented with a new test set containing 1000 phase screens for each of five distinct RMS values ranging from $\lambda/8$ to $\lambda/3$ (5000 total). The average end-of-episode Strehl ratios obtained by the agent for each RMS value are presented in Fig~\ref{fig:rms}.

This experiment is repeated for agents trained with RMS amplitudes of $\lambda/6$ and $\lambda/3$, the results of which are also shown in Fig.~\ref{fig:rms}. Note that all other parameters remain the same as for the reference experiment, with only the scaling factor of the $d_i$ adjusted to allow correction of aberrations up to $\lambda/2$. Notably, this means all agents are trained with an episode length of~4. As shown in \Cref{tab:pv_epsize}, this episode length is insufficient for the agent trained on $\lambda/3$ RMS to reach an acceptable average Strehl ratio after 10M training time steps.

The experiments demonstrate that the agent struggles to generalize beyond the RMS amplitudes it experienced during training. However, it generalizes well to lower amplitudes. Interestingly, the agent trained on $\lambda/3$ RMS, despite underperforming on its training amplitude, achieves comparable results to the other agents on lower amplitudes. This is likely because, within an episode, the agent's actions can influence the residual pupil plane phase, potentially reducing the effective RMS wavefront deviation. Consequently, the agent might have encountered and learned from lower amplitudes during training, even though the original entrance phase screens had a higher RMS value.

These results suggest that for optimal performance, the agent should be trained using the highest RMS amplitude expected in the target application.

\subsection{Impact of noise level on performance}
\label{subsec:noise}

\begin{figure}[htbp!]
\centering\includegraphics[width=\myfraction\linewidth]{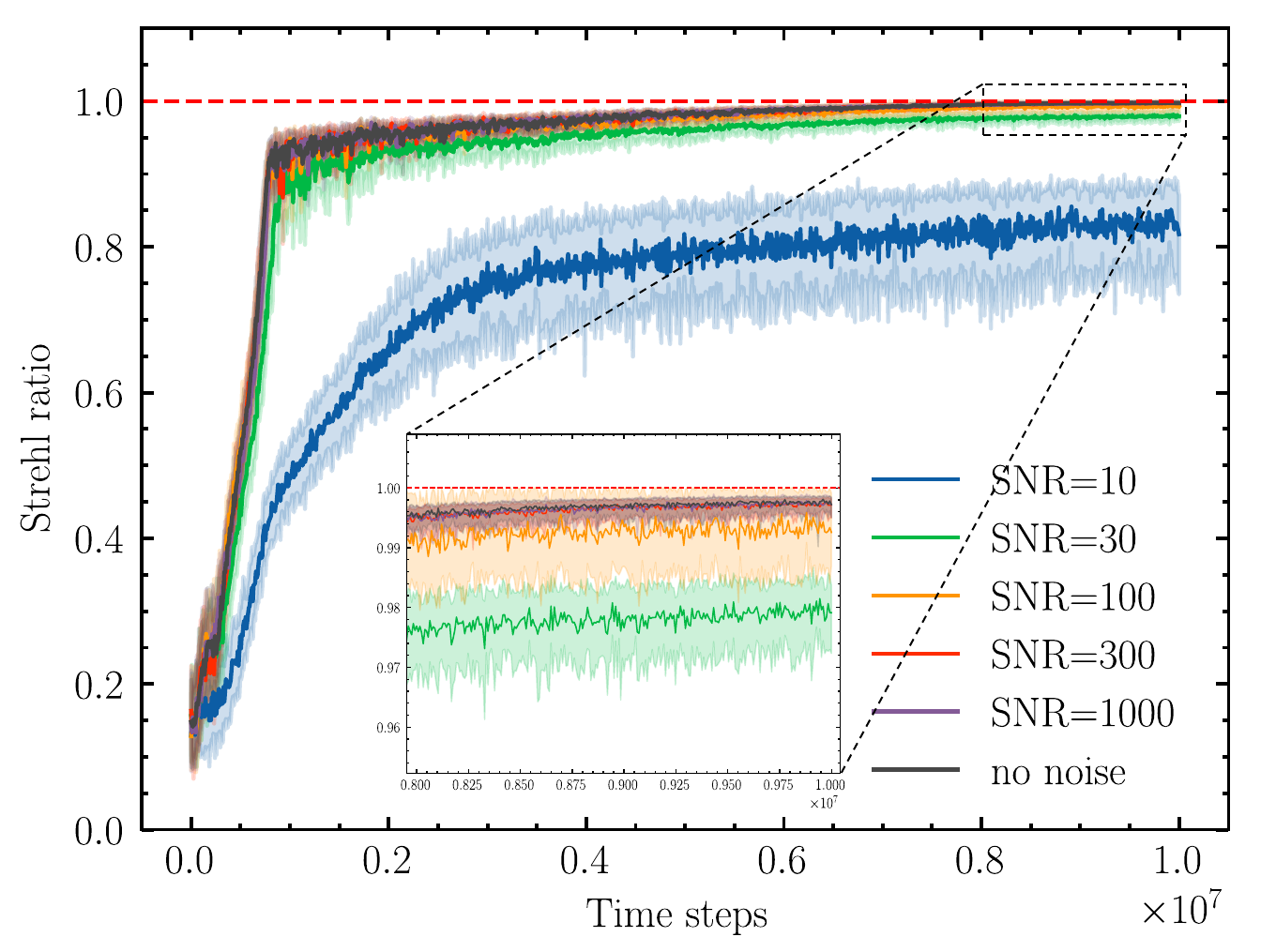}
\vspace*{-.5\baselineskip}
\caption{Learning curves for training SNRs ranging from $10$ to infinity (no noise in the training data), with all other parameters at their nominal value. The error bands represent the spread of the results between the 25th and 75th percentile.}
\label{fig:snr_training}
\end{figure}

\begin{figure}[htbp!]
\centering\includegraphics[width=\myfraction\linewidth]{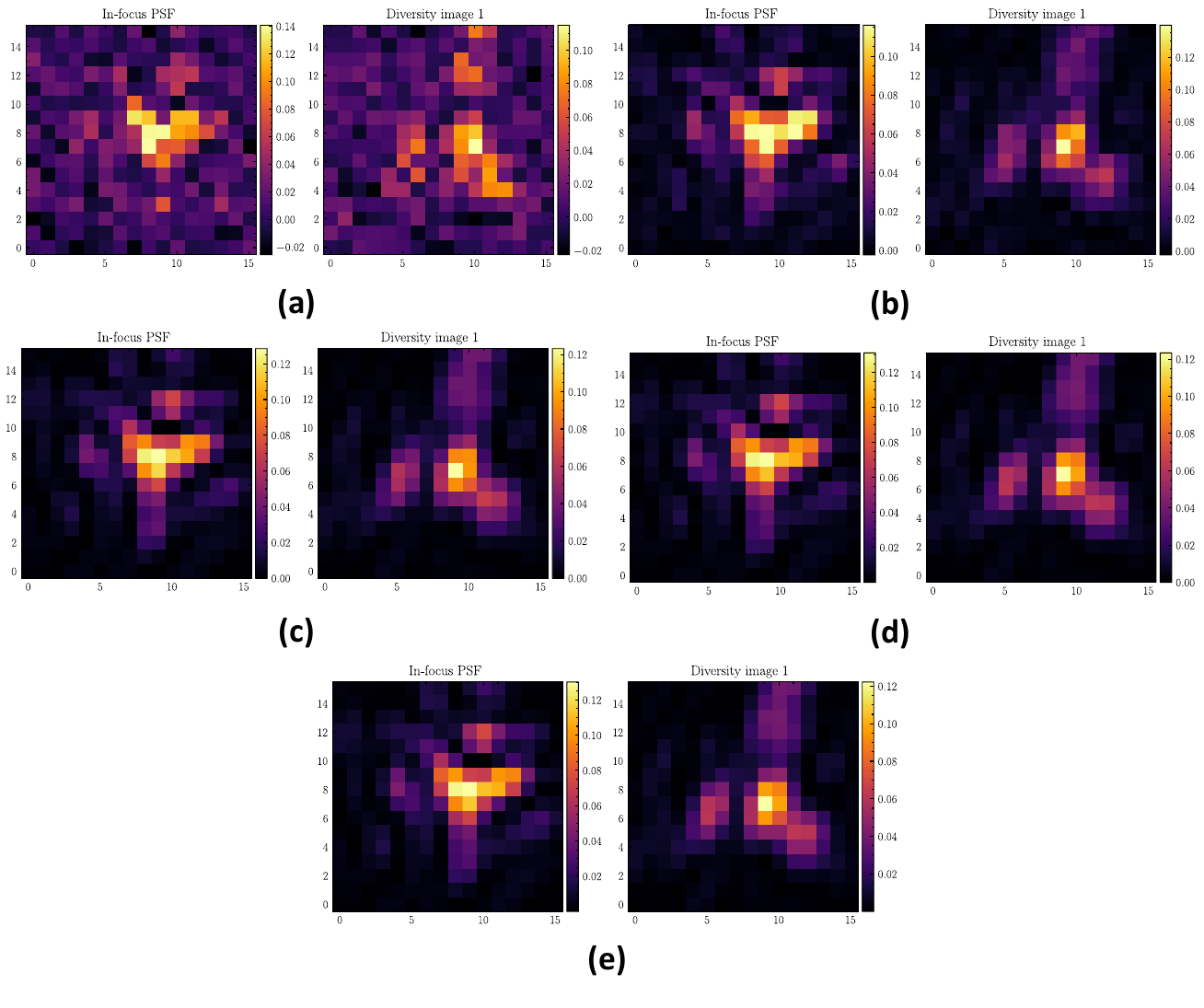}
\vspace*{-.5\baselineskip}
\caption{Examples of observations for a given entrance phase screen, with $SNR = 10$ (a), $30$ (b), $100$ (c), $300$ (d), and $1000$ (e)}
\label{fig:snr_examples}
\end{figure}

This section investigates the method's resilience to noise by examining the agent's performance during training with varying SNR values, ranging from 10 to infinity (no noise), as shown in Fig.~\ref{fig:snr_training}. The benchmark data used for testing is obtained from the same phase screens as the other experiments.
Each curve corresponds to a distinct SNR value, which is identical for both the training data and the benchmark.
In this experiment, the Strehl ratio is obtained before application of the noise, which enables a noise-independent comparison of the curves, as it can be linked to the residual phase after correction (Eq.~\ref{eqn:Mahajan}). 
The plot shows that the agent's performance gradually improves as the SNR increases. While the agent can still achieve some improvement for $SNR=10$, it is barely able to obtain an average Strehl ratio of $0.8$ on the benchmark by the end of training. This is likely because the chosen SNR value reflects the maximum achievable value for $SNR_\text{pix}(x,y)$ (i.e. in the central pixel in the absence of aberrations). However, before correction, the light distribution across the image leads to significantly lower values of $SNR_\text{pix}(x,y)$ in most pixels (see Fig.~\ref{fig:snr_examples}). For example, the typical maximum value of $SNR_\text{pix}(x,y)$ encountered in the observations before correction, is 4 for SNR=10, and 43 for SNR=100. This makes it difficult for the agent to extract meaningful information at high noise levels. Overall, the agent demonstrates a strong capacity to learn correction strategies across various noise levels.

\begin{figure}[htbp!]
\centering\includegraphics[width=\myfraction\linewidth]{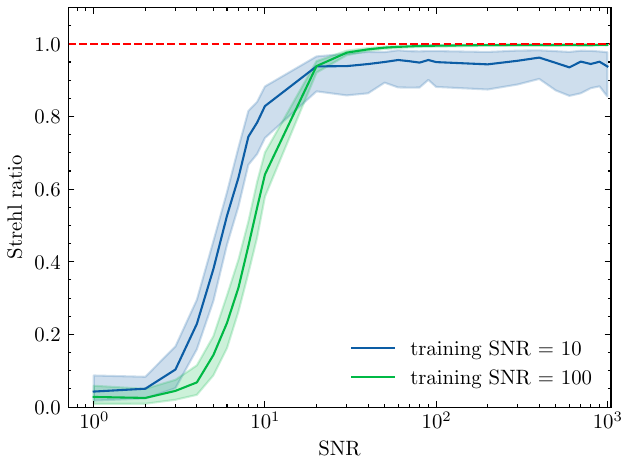}
\vspace*{-.5\baselineskip}
\caption{Average Strehl ratio after correction as a function of the SNR. Two training SNR values (10 and 100) are considered. For each agent, the correction performance is evaluated over 200 test episodes generated with the corresponding SNR. The solid lines represent the average Strehl ratio, and the error bars indicate the range between the 25th and 75th percentile of the results.}
\label{fig:snr_test}
\end{figure}

Fig.~\ref{fig:snr_test} evaluates the agent's generalization capability to unseen noise levels after training completion. We compare the average Strehl ratio after correction as a function of SNR, for two agents trained with a respective SNR of $100$ and $10$. Both agents demonstrate the ability to generalize to SNRs higher than their training values. The agent trained on a higher SNR outperforms the other agent for SNRs above 20. Conversely, the agent trained on a lower SNR exhibits better performance for SNRs below 20, despite obtaining worse results in training. This indicates that training the agent on lower SNR values improves its ability to deal with higher noise levels, albeit at the expense of performance at lower noise levels. Using various SNR values for the observations inside one training may allow the agent to perform well on a wider range of SNRs, and is left to future research.

\subsection{Robustness to uncontrollable modes}
\label{subsec:other}

In this section, we briefly study the scenario where the agent cannot control all the modes present in the aberrations, that is, when $N_\text{modes} > N_\text{act}$. For this experiment, the agent still controls 21 modes, but this time the aberrations contain the first 27 modes (still excluding the piston mode). To maintain consistency with other experiments, the average amplitude of the first 21 aberration modes is set to $125\,\text{nm}$. This results in a total RMS wavefront deviation of approximately $135\,\text{nm}$ for all 27 modes. All other experimental parameters are left unchanged from the reference experiment (in particular, the episode length is left at 4 time steps per episode). 

\begin{figure}[htbp!]
\centering\includegraphics[width=\myfraction\linewidth]{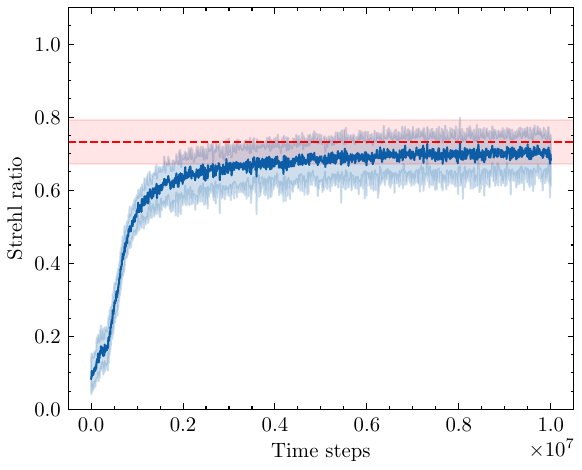}
\vspace*{-.5\baselineskip}
\caption{Learning curve obtained for $N_\text{modes} = 27$ and  $N_\text{act} = 21$, with a total RMS wavefront deviation of $135\,\text{nm}$. All other experimental parameters are left unchanged from the reference experiment. The red dashed line represents the median limit, obtained when the first 21 modes have been suppressed in the benchmark examples. The shaded regions around both the solid and dashed lines show the spread of their respective values between the 25th and 75th percentiles.}
\label{fig:more_modes}
\end{figure}

The learning curve for this experiment is shown in Fig.~\ref{fig:more_modes}. Due to the Zernike polynomials forming an orthogonal basis, there is no way for the agent to influence the 6 extra modes beyond the first 21. Consequently, the maximum attainable Strehl ratio is less than 1. To assess the theoretical performance limit under these constraints, we manually suppress the first 21 modes in each of the 100 benchmark examples. The median Strehl ratio obtained from these modified examples is 0.72 and is represented with a red dashed line in Fig.~\ref{fig:more_modes}. However, depending on the exact value of the 6 uncontrollable modes, this limit might change. To account for this variation, we incorporate a shaded region displaying the range of amplitudes observed in the modified benchmark examples (between the 25th and 75th percentiles of the data). 

Towards the end of training, the agent's average performance almost reaches the average Strehl ratio limit. This shows that the agent is able to learn an effective policy that corrects all the modes that it has control over in spite of the images being degraded by the 6 uncontrollable modes. These modes act as an additional source of noise that the method seems robust to.

\section{Conclusions}

In this work, we investigated the effectiveness of RL for focal plane wavefront control applications. We formulated the focal plane wavefront control problem within the framework of RL, and developed a data-driven method capable of autonomously learning to perform phase retrieval and compute a correction in conjunction. The method uses the PPO algorithm to train an agent to select DM commands using phase diversity images as observations, and a function of the Strehl ratio as reward.

Our results demonstrate that an RL agent can effectively learn to correct up to 36 Zernike modes in various realistic conditions. We determined that episode length is a key parameter to consider when varying the number of modes to correct and the initial aberrations amplitudes. We found that a positive correlation exists between the RMS wavefront error of the training data and the minimum episode length necessary for effective agent training. The method is also robust to SNRs values significantly lower than those encountered during training, and can be extended to noisier images by incorporating observations at lower SNRs in the training data. Finally, preliminary results seem to indicate that the method can learn an effective correction strategy even when the DM cannot fully compensate for the entrance aberrations.

We notice that our method necessitates considerably fewer iterations
than the WFS-less AO techniques discussed in \Cref{subsec:sota}
\cite{ke_self-learning_2019, durech_wavefront_2021,
parvizi_reinforcement_2023}.
At least two novelties contribute to this improvement: the use of a
different reward function, as well as the introduction of the
additional phase diversity image, which provides more information about the state of the environment.
Consequently, we believe that with minor adjustments, our approach could be useful for image-based Adaptive Optics systems using phase diversity. 

While the method demonstrates promising results, there are limitations to address in future work. A topic that requires further investigation is understanding the stability issues occurring when the wavefront deviation exceeds $\lambda/3$ RMS. Additionally, investigating the scalability of the method to high numbers of degrees of freedom (around 1000), using a more realistic DM and working on an actuator basis, will be addressed in future studies.

The relatively large number of training episodes (compared to the methods presented in \Cref{subsec:sota}) required to train the agent to consistently correct the aberrations comes from the fact that PPO, as a model-free algorithm, is quite sample-inefficient. This would likely cause problems when transitioning to experimental data, as training for as many episodes would require an extensive amount of time. This can be addressed by pre-training the agent on simulated data and fine-tuning it on experimental data, following the principles of transfer learning \cite{pan2010transfer}. Therefore, training could primarily rely on simulations with the most accurate model available, utilizing extensive amounts of data. This approach would minimize the need for experimental data, which would only be required in a small quantity towards the end of training to learn the deviations from the model.

 Generating diverse experimental data poses another challenge. In simulations, we can easily generate random, independent aberrations at the start of each training episode. In practice, these could be introduced by a DM. It would be tempting then to use supervised learning instead of model-free RL, since it is more sample-efficient and could accelerate training. However, this approach relies on a model of the DM's response, which can introduce errors in the correction process. Additionally, it would ignore aberrations from other sources, hindering overall performance. Furthermore, using a supervised learning approach would necessitate an extra DM in the optical setup, which could introduce more alignment issues. 
 An alternative approach would be to rely solely on the natural variations of the instrument's intrinsic aberrations. This, however, might limit the agent's exposure to a wide range of phase distortions, potentially hindering its ability to generalize. On the flip side, the agent would encounter similar aberrations more frequently, allowing for more efficient learning on the real-world distribution. Here, transfer learning could also be a solution: the agent could first be trained in simulations on a diverse set of aberrations, and then refine itself on a less varied set of real-world examples that better reflect the experimental distribution. Addressing this question is crucial before proceeding with experimental validation.

Another possible future work is to extend this study to coronagraphic and exoplanet imaging, where the correction precision is currently limited by physical model inaccuracies. This could be addressed by a fully data-driven method such as the one presented in this work.

Future work will involve the implementation of this method on a testbed. Such experimental validation will provide insights into the challenges of transitioning from simulations to experimental data and the applicability of our method to on-sky measurements.

\begin{backmatter}
\bmsection{Acknowledgements}
YG is supported by a Data Intensive Artificial Intelligence (DIAI) PhD scholarship, funded by the Agence Nationale de la Recherche (ANR) and the Data Intelligence Institute of Paris (diiP). OHS, LM and BA acknowledge funding from the \emph{Direction Scientifique Générale de l’ONERA} in the framework of ARE Alioth. LM acknowledges support from project \emph{PEPR Origins}, reference ANR-22-EXOR-0016, supported by the France~2030 plan managed by Agence Nationale de la Recherche. We thank the reviewers for their very constructive criticism and valuable comments that helped improve this paper.

\bmsection{Disclosures} 
The authors declare no conflicts of interest.

\bmsection{Data Availability Statement} 
Data underlying the results presented in this paper were generated on the fly by a code which is not publicly available at this time but may be obtained from the authors upon reasonable request.
\end{backmatter}

\bibliography{Acronymes,EnglishAcronyms,Livres,Articles,yann}

\end{document}